\documentclass[pre,floatfix,twocolumn,showpacs]{revtex4}
\usepackage{amsmath}
\usepackage{amsfonts}
\usepackage{mathrsfs}
\usepackage{epsfig}

\usepackage{graphicx}
\usepackage{dcolumn}
\usepackage{amsmath}

\def\be{\begin{equation}}
\def\ee{\end{equation}}

\begin{document}
%\title{Social polarization as a novel phase of Ising spins on modular networks}
\title{Phase of Ising spins on modular networks analogous to social
polarization}
%\title{Novel phase and slow ordering of Ising spins on modular networks}
%\title{Critical behavior and slow ordering dynamics in modular
%network of Ising spins}
%\title{Extreme slow ordering dynamics in Modular Random Network of Ising
%Spins}
%\title{Slowness of Ordering in Modular Random Network of Ising Spins}
\author{Subinay Dasgupta$^1$, Raj Kumar Pan$^2$ and Sitabhra Sinha$^2$}
\affiliation{%
$^1$Department of Physics, University of Calcutta, 92 Acharya Prafulla 
Chandra Road, Kolkata 700009, India.\\
$^2$The Institute of Mathematical Sciences, CIT Campus, Taramani,
Chennai 600113, India.}
\date{\today}
\begin{abstract}
Coordination processes in complex systems can be related to the problem of
collective ordering in networks, many of which have modular organization.
Investigating the order-disorder transition for Ising spins on modular
random networks, corresponding to consensus formation in society, we
observe two distinct phases: (i) ordering within each module at a critical
temperature, followed by (ii) global ordering at a lower temperature.  This
indicates polarization of society into groups having contrary opinions can
persist indefinitely even when mutual interactions between agents favor
consensus.
%The society is fragmented where there can coexist various opinion among
%people. How such different opinion coexist in society over a period of
%time is often of general interest. We model such a system using modular
%network and by putting Ising spins on the node of the network, we show
%that not only global ordering (or consensus) can be very slow process,
%but under certain condition there can exist a phase such that only the
%module orientation in the same direction is possible and global ordering
%is never achieved.
\end{abstract}

\maketitle

%%%%
\newpage
%%%%
%The nature of transitions in the collective dynamics of interacting
%elements on complex networks have come under scrutiny
%recently~\cite{Dorogovtsev08}.  This is because
Critical phenomena associated with order-disorder transitions is of central
importance in statistical physics~\cite{Dorogovtsev08}.  
%The complex architecture of real-world
%networks may result in very distinct critical behavior for cooperative
%phenomena on such systems as compared to that on regular geometric
%lattices.  
These results also have significant
implications for understanding social phenomena where coordination dynamics
is observed, e.g., in consensus formation~\cite{Castellano00,Lambiotte07}
and adoption of innovations~\cite{Rogers03}.
%, as the links between individuals
%in society often exhibit a complex topology.  
Such processes can be analyzed using the spin models of statistical physics
which are generic systems for studying cooperative phenomena.  The
spin orientations correspond to several equivalent but mutually exclusive
choices made by an agent on the basis of information about the choice of
the majority in its local neighborhood. The simplest case is when an agent
decides between two competing choices where the dynamics can be modeled by
an Ising system defined on a network reflecting the social contact pattern.
%{\bf We may introduce the existence of different opinion in this paragraph.
%In the next paragraph we show how it is possible using the model.}

One of the most remarkable features of complex networks seen in 
many different contexts is their modular
organization~\cite{Girvan02}. Modular networks
consist of subnetworks whose nodes have a significantly higher connection
density compared to the overall density of the network
\cite{Pan07,Pan08,Pan09a}. A recent analysis of a network of mobile phone
users, reconstructed according to their call frequency and duration, have
shown the existence of such modular structure in the organization of social
contacts \cite{Onnela07}.  This has also been observed in other social
networks, e.g., of scientific collaborators~\cite{Palla05}, electronic mail
communication \cite{Guimera03,Tyler05}, the PRETTY-GOOD-PRIVACY(PGP) encryption
``web-of-trust''~\cite{Boguna04} and even that of non-human
animals~\cite{Lusseau04}.  Study of dynamical processes on such
networks~\cite{Strogatz01} can help in understanding how individual
behavior at the microscopic level relates to social phenomena at the
macroscopic level.

\begin{figure}
\begin{center}
  \includegraphics[width=0.8\linewidth]{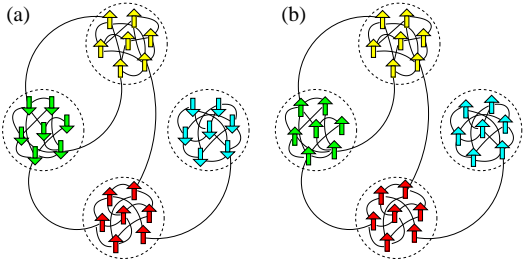}
  \end{center}
	\caption{(Color online) Ordering in modular networks: (a) Local ordering within modules
	without global order, and (b) the globally ordered state.} 
\label{fig:spin_model}
\end{figure}

In this Rapid Communication, we investigate a modular network of Ising
spins with ferromagnetic interactions, and report the existence of a 
phase with ordering among spins in each module,
in the absence of global ordering. This modular ordered phase
(Fig.~\ref{fig:spin_model},~a) is reached from the disordered state of the
system through a continuous transition by lowering the temperature to the
critical value $T_c^m$. Further reducing the temperature to $T_c^g$ results
in another continuous transition to a state where all the modules are
aligned in parallel, i.e., the globally ordered phase
(Fig.~\ref{fig:spin_model},~b).  
%The two critical temperatures converge as
%the network becomes more homogeneous with decreasing modularity.
%As the modularity of the network is decreased, the critical temperatures
%for appearance of modular and global ordering converge, such that in a
%relatively homogeneous network the mean-field result for the Ising model
%is recovered. 
Our results help to understand how contrary opinions can co-exist in
society even when mutual interactions favor consensus.
In the social context, the concept of physical temperature corresponds to
a measure of noise, e.g., due to imperfect
information or uncertainty on the part of the agent, or the existence of
idiosyncratic beliefs.  
Below the critical temperature $T_c^g$, the relaxation time to global order
diverges as the modularity increases.  
%results imply that a social network can have clusters Even when global
%order is possible, if the modules are sparsely connected, the ordering
%dynamics can take extremely long so that ordering may not be observed
%within a reasonable period.  This is possibly the reason for recent
%reports of the existence of a phase without global ordering at low
%T~\cite{Suchecki06}.  
Thus, even when global order is achievable (i.e., $T < T_c^g$), the time
required to achieve consensus increases rapidly as the social organization
becomes more modular.  However, this time can be altered significantly by
considering processes that are observed in society. We implement this by
(i) introducing an external
field modeling positive feedback effects which reinforce the choice adopted by
the majority~\cite{Arthur89}, or
(ii) varying the strength of couplings between communities relative to those
of intra-modular links, which is consistent with the well-known ``weak tie'' 
hypothesis for social networks~\cite{Granovetter73}.

We consider a modular random network consisting of $N$ nodes arranged 
into $n_m$
modules, each having $n$ ($= N/n_m$) nodes~\cite{Pan09a}.  
The connection probability between a pair of nodes
belonging to the same module is $\rho_i$, while that for nodes belonging to
different modules is $\rho_o$.  The modularity of the network can be
changed continuously by varying the ratio of inter-modular to intra-modular
connectivity, $r = \frac{\rho_o}{\rho_i} \in [0,1]$, keeping the average
degree $\langle k \rangle$ constant.  For $r = 0$, the network
gets fragmented into a set of isolated clusters, while at $r = 1$, 
it becomes a homogeneous or Erdos-Renyi random network. On placing
Ising spins on the nodes of the network, the Hamiltonian for the system is
given by
\begin{equation}
H = - \sum_{i,j} J_{ij} \sigma_i \sigma_j,
\label{hamiltonian}
\end{equation}
where $\sigma_i = \pm 1$ is the Ising spin on the $i$th node and the
coupling strength $J_{ij}$ is $J (>0)$ if $i, j$ are connected, and zero
otherwise. The positive value of $J$ (ferromagnetic coupling) implies that
each link tries to align the two spins connected by it.
%Fig.~\ref{fig:spin_model} shows the two possible scenarios of spin
%ordering in such a system, viz., global order, in which all spins are
%aligned in parallel, and modular order, in which although the spins within
%each module are aligned parallely, different modules have different
%orientations resulting in an overall low magnetization.

\begin{figure}
\begin{center}
  \includegraphics[width=0.95\linewidth]{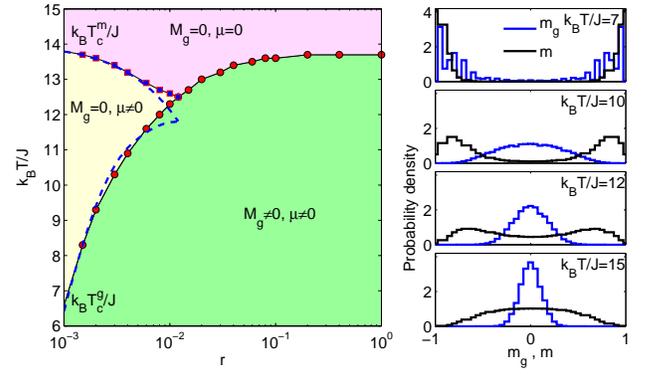}
  \end{center}
	\caption{(Color online) Critical behavior of Ising model on modular network. (a) The
	$r$-$k_{B}T/J$ phase diagram indicating the existence of three distinct phases:
	disordered ($M_g = \mu = 0$), modular order ($M_g = 0$, $\mu \neq 0$) and
global order ($M_g \neq 0$, $\mu \neq 0$).  The solid lines are phase
boundaries obtained from Monte Carlo simulation,
while the broken lines indicate the critical temperatures $T_c^m$ and
$T_c^g$ obtained from analytical expressions
given in the text.  (b) The distributions for global ($m_g = M_g/ N$) and
modular ($m = \mu / n$) magnetizations per site at different temperatures
showing the two phase transitions for $r = 0.002$.
%The different region in the $r$-$T$ plane shows the existence of
%different phases. If we only look at relaxation behavior upto some
%finite time, say $\tau_0=10^{-5}$, a virtual phase may appear. However
%this is not a true phase and disappears in the limit $t \rightarrow
%\infty$.  We also look at distribution of global and modular
%magnetization at various temperature. When temperature is low, say
%$T=8$, both the $M$ and $M_m$ 
At low temperature ($T = 7 J/k_B$) both curves have bimodal nature,
indicating global order. As $T$ increases to $10 J/k_B$ ($> T_c^g$),
the global order disappears resulting in unimodal distribution of $m_g$.
For temperatures higher than $T_c^m$ ($T = 15 J/k_B$), both curves
are unimodal as the system is disordered.  All results shown are for $N =
512$, $n_m = 16$ and $\langle k \rangle$ = 14.} 
\label{fig:phase}
\end{figure}

Starting from a disordered state, we have performed Monte Carlo simulation
for updating the spin states using the Wolff algorithm~\cite{Newman99}.
For any configuration, we can measure the average magnetic moment per
module, $\mu = \langle |\sum_{i \in s} \sigma_i| \rangle$, the
averaging being over all modules $s = 1, \ldots, n_m$, and, 
the total (or global) magnetization
of the network, $M_g = \sum_i \sigma_i$.  At equilibrium, the resulting
phase diagram (Fig.~\ref{fig:phase}) clearly shows the existence of three
phases, corresponding to (a) the disordered state ($\mu = 0, M_g = 0$), (b)
a globally ordered state ($\mu \neq 0, M_g \neq 0$) and (c) a state with
only modular order ($\mu \neq 0, M_g = 0$).  At low $r$, as the temperature is
decreased the system undergoes two transitions: first, at $T=T_c^m$, from
the disordered state to the state with modular order, and next, at
$T=T_c^g$, to the globally ordered state, where all spins in the network
are aligned in parallel.  As the modularity of the network decreases with
increasing $r$, the two critical temperatures $T_c^m$ and $T_c^g$ approach
each other and finally converge.

The phase diagram can be reproduced analytically by considering the free energy
for the system. The magnetic moment of a single module is $\mu = n (2 f_+
-1)$, where $f_+$ is the fraction of ``up'' ($\sigma = +1$) spins in that
module. We assume that, at equilibrium, the magnetic moment of each of the
$n_m$ modules has the same magnitude, with a fraction $f_+^{m}$ being
positive ($+ \mu$) and the remainder being negative ($- \mu$).
This is valid in the regime of {\em strong modularity}, i.e., $r \ll 1$.
Thus, the total magnetic moment for the system is $M_g = n_m \mu (2 f_+^{m}
- 1)$.  The contribution to the internal energy of the system from each
module is $U_i = - J L_i (2 f_+ - 1)^2$, where $L_i = \frac{n(n-1)}{2}
\rho_i$ is the number of links within a module. This is based on the
mean-field assumption that the neighborhood of all spins are identical. To
obtain the internal energy contribution for interactions between modules,
$U_o$, we note that each of the modules can be treated as ``spins'' of
moment $\mu$ with $L_o = \frac{n_m (n_m - 1)}{2} \rho_o$ links between
them. Analogous to the preceding expression for $U_i$, we get 
$U_o = - J L_o \mu^2 (2 f_+^{m} - 1)^2$.  
Then, the free energy for the system is
\begin{eqnarray}
F (f_+, f_+^{m}) = n_m (U_i - T S_i) + U_o - T S_o,
%F (f_+, f_+^{m}) = U_o + n_m U_i
%\nonumber \\
%\nonumber
% + n_m k_B T \left[ f_+^{m} \log (f_+^{m}) + 
%(1 - f_+^{m}) \log (1 - f_+^{m}) \right] \\
% + n n_m k_B T \left[ f_+ \log (f_+) + 
%(1 - f_+)\log (1 - f_+) \right],
\label{free_energy}
\end{eqnarray}
where the entropy terms, $S_i = - n k_B [f_+ \log (f_+) +(1 - f_+)\log (1 -
f_+)]$, and $S_o = - n_m k_B [f_+^{m} \log (f_+^{m}) + (1 - f_+^{m}) \log
(1 - f_+^{m})]$, correspond respectively to the different ways in which $n
f_+$ up spins can be distributed among $n$ spins within each module,
and $n_m f_+^{m}$ modules with moment $+\mu$ can be distributed among $n_m$
modules.

%For $r < r_c$ and finite $n$, minimizing the free energy $F$
%(Eq.~\ref{free_energy}) with respect to $f_+$ we get,
To analyze the critical behavior of the system we minimize the free
energy $F$ (Eq.~\ref{free_energy}) with respect to $f_+$,
giving:
\begin{equation}
\frac{1}{4f_+ - 2}\log \frac{f_+}{1 - f_+} =  \frac{T_c^m}{T}
+  \frac{J \rho_o n(n_m - 1)(2f_+^{m} - 1)^2}{k_B},
\label{T_c_1}
\end{equation}
where, $T_c^m = 2 J L_i/(n k_B)$. 
This indicates a continuous phase
transition at the modular critical temperature $T_c^m$, 
below which spins within a module are ordered
but $f_+^m = 1/2$.  This state corresponds to the phase for which
$\mu \neq 0, M_g = 0$.  
As the temperature decreases below $T_c^m$, the different modules get
aligned with each other at a temperature $T_c^g$. This is obtained by
minimizing $F$ with respect to $f_+^{m}$:
\begin{equation}
\frac{1}{2(2f_+^{m} - 1)}\log \frac{f_+^{m}}{1 - f_+^{m}} =  \frac{T_c^g}{T},
\label{T_c_2}
\end{equation}
which shows a continuous phase transition at the global critical temperature
$T_c^g = \frac{J \rho_0}{k_B}(n_m - 1) n^2 (2f_+ - 1)^2$.
The expression for $T_c^g$ does not have a closed analytic form and it 
is obtained by numerical minimization. 
As $r \rightarrow 1$, the network loses its modular structure and becomes
a homogeneous random network, so that Eqs.~\ref{T_c_1}-\ref{T_c_2} give 
the critical temperature $T_c^m = T_c^g = J \langle k \rangle / k_B$.
%effectively $n_m = 1$.

%For $r > r_c$, near the critical temperature there are large fluctuations
%in $\mu$ and the entropy $S_o$ can be neglected. Thus, the free energy
%expression of Eq.~\ref{free_energy} simplifies to,
%\begin{equation}
%F (f_+, f_+^{m}) =n n_m (2 f_+ -1)^2 J \langle k \rangle /2 - n_m T S_i. 
%\label{free_energy2}
%\end{equation}
%Minimizing this expression with respect to $f_+$ we get
%\begin{equation}
%\frac{1}{4f_+ - 2}\log \frac{f_+}{1 - f_+} =  \frac{J\langle k \rangle}{k_BT}.
%\label{T_c_large}
%\end{equation}
%This equation will have a solution $f_+ > 1/2$ only when $k_B T < J\langle
%k \rangle$ indicating that there is a continuous phase transition at the
%critical temperature $T_c^m = T_c^g = J\langle k \rangle / k_B$ which
%agrees with the numerical results (Fig.~\ref{fig:phase}).{\bf I have my
%doubts over this paragraph: (1) When the modularity is lost it is obvious
%that the $T_c=J\langle k \rangle /k_B$ because it is the mean field limit
%of a random network.  (2) The comment that it agrees with the numerical
%results is also not true. Because the numeric suggest that $T_c^g$ varies
%smoothly with $r$ finally converging to the above value only at r=1. So
%just putting a line for $T_c^g$ in the phase diagram, which in this case is
%14 will only indicate that this is true only near $r=1$.}

The above analysis for finite-size networks are valid even in the {\em
thermodynamic limit} when it is approached by increasing the number of
modules, $n_m$. Note that, if we instead increased the number of nodes in a
module, $n$ (keeping $n_m$ fixed), the modularity of the network is lost
and hence, there is only a single continuous order-disorder transition at
$T_c = J \langle k \rangle / k_B$.  The two types of ordering seen for the
Ising model on a modular network imply that not only can consensus formation
in a society be an extremely slow process~\cite{Castello07}, 
but under certain conditions it
may never be achieved and communities with contrary opinions can persist
indefinitely.  

\begin{figure}
\begin{center}
  \includegraphics[width=0.8\linewidth]{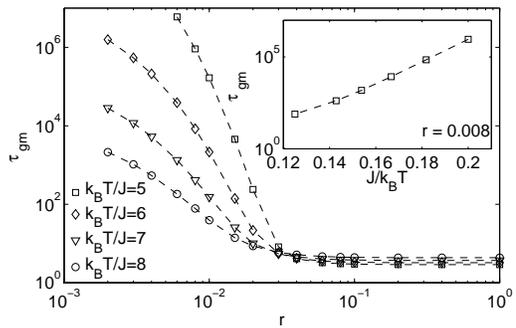}
  \end{center}
  \caption{Relaxation time to the globally ordered state, $\tau_{gm}$, as 
  a function of $r$ at different temperatures. With increasing modularity,
  $\tau_{gm}$ diverges. (Inset) The variation in $\tau_{gm}$ with $1/T$ 
  at $r=0.008$. For all cases, $N = 512$, $n_m = 16$ and $\langle k \rangle = 14$.}
	\label{fig:tau_gm}
\end{figure}

We now look at how the time required to reach the equilibrium state
corresponding to global order varies with the
modularity of the network. Starting from an initially disordered state
($M_g=0$, $\mu=0$), spins are updated using Glauber dynamics~\cite{Glauber63}.
%which
%involves picking a spin at random and flipping it with probability $p = 1/
%(1 + \exp (\delta E /k_B T)$, where $\delta E$ is the change in energy
%resulting from the flip. {\bf Is the expression for Glauber dynamics
%needed, people often write the magnetization equation directly after citing
%R.~J. Glauber, J. Math. Phys. {\bf 4} 294 (1963)}
%where 1 Monte Carlo (MC) step
%corresponds to: (a) a spin $i$ being chosen randomly and its energy $E_i$
%calculated from Eq.~\ref{hamiltonian}, (b) the spin is flipped with
%probability $\frac{\exp (\beta E_i)}{\exp (\beta E_i) + \exp (- \beta
%E_i)}$ ($\beta = 1/k_B T$), (c) repeat $N$ times steps (a) and (b).
Fig.~\ref{fig:tau_gm} shows that the relaxation to global order takes
an extremely long time as the system become more modular.
To understand this, we first consider a single {\em isolated} module of $n$ nodes with $L_i$ intra-modular links. 
Under the mean-field approximation, the time-evolution of the magnetization
per site $m = \mu / n$ is described by 
\begin{equation}
\frac{dm(t)}{dt} + m (t) = \tanh \left[ \frac{T_c}{T} m (t) \right], 
\label{mag_evoln}
\end{equation}
where $T_c = 2JL_i/(n k_B)$.  At equilibrium, this reduces to the usual
mean-field equation, $m = \tanh (m T_c/T)$, indicating that $T_c$ is the
critical temperature.
%Solving Eqn.~\ref{mag_evoln}, we find the time required to attain
%magnetization $m$: $t = \int_{0^+}^m \frac{dx}{\tanh(xT_c/T) -
%x}$~\cite{note2}.
Defining {\em relaxation time} $\tau$ to be the time in which $m$ becomes
0.5, we obtain from Eq.~\ref{mag_evoln},
\begin{equation}
\tau = \int_{0^+}^{0.5} \frac{dx}{\tanh(xT_c/T) - x}.
\label{mag_tau}
\end{equation}
%$~\cite{note2}.
Although it is small at low temperatures, $\tau$ diverges as $T \rightarrow
T_c$ due to critical slowing down.

In a network consisting of several modules, the relaxation time to modular
order, $\tau_{mm}$, is identical to $\tau$ derived above with $T_c =
T_c^m$.  To obtain the relaxation time to global order, we first note that
the free energy of a single module $F_m(f_+) = U_i - T S_i,$ 
%= - J L_i (2f_+ - 1)^2 + n k_B T [f_+\log f_+ + (1-f_+)\log(1-f_+)]$
has two minima at $f_+^{0}$ and $1-f_+^{0}$ (say). These correspond to the
magnetic moments $\pm \mu$, which are separated by a free energy maximum
at $f_+ = 1/2$ corresponding to $\mu=0$.  To switch
from $+\mu$ to $-\mu$ (or vice versa), a module has to overcome an energy
barrier $ \Delta = F_m(1/2) - F_m(f_+^{0})$.  Thus, the time required to
attain a global magnetization $M_g$ is slowed down by the factor
$\exp[\Delta/(k_B T)]$.  Defining the global relaxation time, $\tau_{gm}$,
to be the time in which the global magnetization per site, $M_g/(n_m\mu)$ 
becomes 0.5, we obtain from Eq.~\ref{mag_tau},
%where $T_c^{2}$ is given by Eq. (11).  If we define, as before, the
%relaxation time $\tau_{gm}$ as the one for which the quantity $(M_g/n_m
%n)$ attains the value 0.5, then the expression for it will be
\begin{equation}
\tau_{gm} = \exp\left(\frac{\Delta}{k_B T}\right)\int_{0^+}^{0.5/|2f_{+}^{0} -1|} 
\frac{dx}{\tanh(xT _c^g/T) - x}. 
\label{tau_global}
\end{equation}
Here we have assumed that $\tau_{gm} \gg \tau_{mm}$, 
which is valid
for $T \ll T_c^{g}$ and low $r$.  In this region, $\tau_{gm}$ diverges
with increasing modularity of the network, the trend agreeing with the
numerical results of Fig.~\ref{fig:tau_gm}. Note
that, at low temperatures ($T \ll T_c^g$), the integral in
Eq.~\ref{tau_global} can be neglected and $\tau_{gm} \approx
\exp(\Delta/k_B T) \propto \exp (J\rho_i/k_B T)$, assuming $f_+^{0} \approx
1$. Thus, the relaxation to global order becomes exponentially 
slower as the temperature
decreases (Fig.~\ref{fig:tau_gm}, inset). The above results indicate that
even when $T < T_c^g$, 
a strongly modular network may require an extremely long time to reach the
globally ordered state, which explains why previous
studies may have erroneously observed a separate phase without global order
in this region~\cite{Suchecki06}. 

\begin{figure}
\begin{center}
  \includegraphics[width=0.98\linewidth]{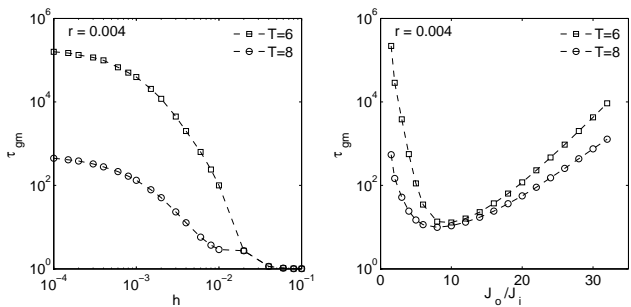}
  \end{center}
  \caption{The relaxation time to the globally ordered state, $\tau_{gm}$, as a function of
  (a) the external field $h$, and, (b) the ratio of inter to intra-modular 
  coupling strength $J_o/J_i$, at different temperatures. The minima of 
  $\tau_{gm}$ occur when $J_o/J_i \simeq \langle k_{in} \rangle$, the
  average number of links a node has with other nodes in its module.
  For all cases, $N = 512$, $n_m = 16$, $\langle k \rangle = 14$ 
  and $r = 0.004$.}
	\label{fig:field}
\end{figure}

It may appear from the preceding analysis that achieving global consensus
is extremely difficult in a real social network having modular
organization. However, we now discuss possible mechanisms by which the time
to attain global order can be changed significantly. First, we look at the
role of an external magnetic field which is proportional to the global
magnetization $M_g$. This corresponds to positive feedback effects in
social systems, where, although two competing alternatives are initially
equivalent, as more and more agents switch to one particular option, it
becomes the preferred choice~\cite{Arthur89}. 
%An example is the competition between two computer operating systems, which are
%initially having the same capabilities.  However, as third party providers
%are more interested in developing new software for the system that the
%majority uses, when small fluctuations drive the user base of one system
%higher, the latter has an advantage in attracting more users as there is
%more choice in terms of the programs available for the system.  Thus, a
%positive feedback effect sets in to eventually make one choice
%overwhelmingly superior.  
Introducing such a field, the Hamiltonian for the system becomes $H =
\sum_{i,j} J_{ij} \sigma_i \sigma_j - h (\sum_i \sigma_i)^2$.
The external field $h$ has no effect in the
absence of global order, but when $M_g$ is non-zero, the field drives the
system to the equilibrium state corresponding to global order much faster
as seen in 
Fig.~\ref{fig:field}~(a). 
The free energy in this case can be obtained by
replacing $J L_m$ with $J L_m + h n_m^2$ in Eq.~\ref{free_energy}. 
Thus, the field effectively increases the inter-modular
interactions (making the network less modular) which drives the system away
from the critical point by increasing $T_c^g$, thereby reducing $\tau_{gm}$.

We next consider the possibility that the interaction strengths for
inter-modular connections ($J_o$) may be different from those of
intra-modular links ($J_i$). From Eq.~\ref{free_energy}, it is clear that
increasing the ratio $J_o/J_i$ is equivalent to increasing $r$.  Therefore,
as the intra-modular links increase in strength relative to the
inter-modular links (making the network even more modular), the time to
achieve global order ($\tau_{gm}$) increases. On the other hand, when the
inter-modular links are stronger, $\tau_{gm}$ decreases up to a point and
then, with increasing $J_o$, starts increasing as the nodes in different
communities get strongly coupled thereby destroying the identity of
individual modules (Fig.~\ref{fig:field},~b).  This {\em non-monotonic} behavior
of relaxation time as a function of the ratio of strengths for short- and
long-range interactions is in contrast with that seen in the case of
Watts-Strogatz (WS) small-world network model~\cite{Jeong05}.  This
reinforces previous results that the dynamics of WS networks are strikingly
different from that of modular networks, although their structural
properties are similar~\cite{Pan09a}. It also suggests that a system may
maintain diversity by using weak links between their constituent
communities~\cite{Granovetter73}. 

In this Rapid Communication we have shown that the order-disorder transition in an Ising
model defined on a modular network shows the existence of three distinct
phases.  While the disordered and globally ordered phases are similar to
those expected for a homogeneous network, the existence of a phase with
local ordering within modules but no global order is a novel effect of the
modular structure.  This has significant implications for ordering dynamics
in real systems, e.g., consensus formation on social networks. It indicates
that, under certain conditions, homogeneous groups with contrary opinions
can coexist forever even when mutual interactions between agents favor
consensus, so that society becomes polarized.  Our results suggest that
such tendencies can be countered by increasing inter-community
communication and improving the overall penetration of mass media in the
society.
%Even for the parameter region where
%global order is possible, the time to reach such an equilibrium
%can be extremely long for strongly modular systems.
%, in agreement with previous studies of related models~\cite{Castello07}. 
%reported different timescales of ordering~\cite{Hinczewski07}, that
%may result in a long-lived disordered state with communities having
%different opinions~\cite{Castello07}, in our system global order will
%forever be absent under certain conditions. 
Although the present Rapid Communication looks at the case of two competing choices, it
is possible that the effects seen here extend to the case of agents
choosing between multiple ($> 2$) alternatives, e.g., in the context of
$q-$state Potts spin dynamics on modular networks. Further, the network can
be made more realistic from a social perspective by considering the
presence of hubs and a broad link-strength distribution~\cite{Onnela07}.
In summary, our results imply that
due to mesoscopic structural inhomogeneity, equilibrium as well as
dynamical properties of local regions in a complex network may depart
significantly from those for the entire network. 

We thank P.~M. Gade and P.~Ray for helpful discussions. This work was
supported in part by CSIR, UGC-UPE and IMSc Complex Systems (XI Plan)
Project. 

%{2}

\end{document}